\documentclass[sigconf]{acmart}

\usepackage{subfig}
\AtBeginDocument{%
  }

\copyrightyear{2025}
\acmYear{2025}
\setcopyright{rightsretained}
\acmConference[WWW Companion '25] {Companion Proceedings of the ACM Web Conference 2025}{April 28-May 2, 2025}{Sydney, NSW, Australia.}
\acmBooktitle{Companion Proceedings of the ACM Web Conference 2025 (WWW Companion '25), April 28-May 2, 2025, Sydney, NSW, Australia}
\acmISBN{979-8-4007-1331-6/25/04}
\acmDOI{10.1145/XXXXXX.XXXXXX}
\begin{document}

\title{Pre-train and Fine-tune: Recommenders as Large Models}

\author{Zhenhao Jiang}
\authornote{Both authors contributed equally to this research.\\This work was done during Zhenhao Jiang's internship at Alibaba Group.}
\email{222041010@link.cuhk.edu.cn}
\orcid{1234-5678-9012}
\affiliation{%
  \institution{The Chinese University of Hongkong, Shenzhen}
  \city{Shenzhen}
  \state{Guangdong}
  \country{China}
}

\author{Chenghao Chen}
\authornotemark[1]
\email{chenchenghao.cch@alibaba-inc.com}
\affiliation{%
  \institution{Alibaba Group}
  \city{Hangzhou}
  \state{Zhejiang}
  \country{China}
}

\author{Hao Feng}
\email{zhisu.fh@alibaba-inc.com}
\affiliation{%
  \institution{Alibaba Group}
  \city{Hangzhou}
  \state{Zhejiang}
  \country{China}
}

\author{Yu Yang}
\email{huke.yy@alibaba-inc.com}
\affiliation{%
  \institution{Alibaba Group}
  \city{Hangzhou}
  \state{Zhejiang}
  \country{China}
}

\author{Jin Liu}
\email{nanjia.lj@alibaba-inc.com}
\affiliation{%
  \institution{Alibaba Group}
  \city{Hangzhou}
  \state{Zhejiang}
  \country{China}
}

\author{Jie Zhang}
\email{shenxu.zj@alibaba-inc.com}
\affiliation{%
  \institution{Alibaba Group}
  \city{Hangzhou}
  \state{Zhejiang}
  \country{China}
}

\author{Jia Jia}
\email{jj229618@alibaba-inc.com}
\affiliation{%
  \institution{Alibaba Group}
  \city{Hangzhou}
  \state{Zhejiang}
  \country{China}
}

\author{Ning Hu}
\affiliation{%
  \institution{Alibaba Group}
  \city{Hangzhou}
  \state{Zhejiang}
  \country{China}
}

\renewcommand{\shortauthors}{Zhenhao Jiang et al.}

\begin{abstract}
  In reality, users have different interests in different periods, regions, scenes, \textit{etc}. Such changes in interest are so drastic that they are difficult to be captured by recommenders. Existing multi-domain learning can alleviate this problem. However, the structure of the industrial recommendation system is complex, the amount of data is huge, and the training cost is extremely high, so it is difficult to modify the structure of the industrial recommender and re-train it. To fill this gap, we consider recommenders as large pre-trained models and fine-tune them. We first propose the theory of the information bottleneck for fine-tuning and present an explanation for the fine-tuning technique in recommenders. To tailor for recommendation, we design an information-aware adaptive kernel (IAK) technique to fine-tune the pre-trained recommender. Specifically, we define fine-tuning as two phases: knowledge compression and knowledge matching and let the training stage of IAK explicitly approximate these two phases. Our proposed approach designed from the essence of fine-tuning is well interpretable. Extensive online and offline experiments show the superiority of our proposed method. Besides, we also share unique and important lessons we learned when deploying the method in a large-scale online platform. We also present the potential issues of fine-tuning techniques in recommendation systems and the corresponding solutions. The recommender with IAK technique has been deployed on the homepage of a billion-scale online food platform for several months and has yielded considerable profits in our business.
\end{abstract}

\begin{CCSXML}
<ccs2012>
   <concept>
       <concept_id>10002951.10003317.10003347.10003350</concept_id>
       <concept_desc>Information systems~Recommender systems</concept_desc>
       <concept_significance>500</concept_significance>
       </concept>
   <concept>
       <concept_id>10002950.10003712</concept_id>
       <concept_desc>Mathematics of computing~Information theory</concept_desc>
       <concept_significance>500</concept_significance>
       </concept>
 </ccs2012>
\end{CCSXML}

\ccsdesc[500]{Information systems~Recommender systems}
\ccsdesc[500]{Mathematics of computing~Information theory}

\keywords{Recommendation System, Information Bottleneck, Fine-tuning Technique}

\received{20 February 2007}
\received[revised]{12 March 2009}
\received[accepted]{5 June 2009}

\maketitle

\section{Introduction}
Shipping billions of items to meet users' tastes is the basic functionality of recommendation systems \cite{lu2012recommender, lu2015recommender, liu2023chatgpt}, which is an important part of online information distribution platforms such as news feeds, e-commerce, social media, \textit{etc.} At present, the mainstream recommender systems, based on deep learning \cite{he2016deep, jiang2023mobile}, use an end-to-end training strategy to mine user interest and rank items according to user preference \cite{chen2021end}. With the development of business, there are more and more topics/conditions in the domain of recommender algorithms. Multi-scene \cite{tan2021multi, he2020dadnn}, multi-region \cite{oh2017real, gomez2022provider}, and multi-period \cite{ahmadian2022alleviating, singh2022novel} are three representatives becoming the research frontier. Multi-scene describes that online recommender service consists of many scenes, such as home page, channel page, in-shop related page, \textit{etc}. Users may click back and forth on different scenes with distinct behaviors. Therefore, it is necessary to explicitly model user-item interaction behavior in different scenes. Multi-region mainly includes multi-country and multi-city, which describes the differences in life habits and preferences of users among different regions, as well as the differences in the supply of items. For example, traditionally, turkey is for Thanksgiving in America while zongzi is for the Dragon Boat Festival in China. Multi-period describes that users have different behaviors at different times. For example, morning \textit{vs.} evening on takeaway platforms, and holiday \textit{vs.} workday on short video platforms. Generally speaking, we can consider multi-scene, multi-region, and multi-period as belonging to the scope of multi-domain learning.

 There are several early studies \cite{ning2023multi, zhang2023meta} that address the problems with the multi-domain learning paradigm that employs one model to serve all domains. However, if a new task is added, the entire model needs to be retrained, which is very inconvenient and increases the training overhead. \textbf{Remark:} Large-scale online platforms generate data volumes of several terabytes (Tb) per day, and recommender algorithms are incrementally learning with newly generated data at regular intervals rather than training from scratch. Therefore, when the model structure is massively modified, it means that a large number of parameters need to be trained from scratch, and more historical data (thousands of Tb) may be needed to train the newly modified structure, which greatly increases the training cost that is unacceptable in industrial recommender. 

Under different conditions/domains, users' interests and behaviors change. Therefore, the recommendation tasks under different conditions/domains can be regarded as downstream tasks, and the pre-training and fine-tuning methods can be used to conduct more tailored modeling. Compared to multi-domain learning, fine-tuning requires only minor changes to the original model. Adding a new task also only requires fine-tuning on that task and has no effect on the original model \cite{hu2021lora}. Therefore, we consider treating our recommender as a large model and fine-tuning it on different domains.

Recently, large language models (LLMs) \cite{christiano2017deep, biswas2023potential} have taken the world by storm for their amazing performance in the field of natural language processing (NLP). Similarly, taking the multi-period task in a takeaway platform as an instance, the large-scale industrial recommender is trained on the full dataset covering all time, which learns the general business knowledge and the common preferences of users. However, due to the huge natural gap between different periods (\textit{e.g.} breakfast \textit{vs.} dinner), it is difficult for the model to finely trace the shift in user preference at different periods. Fortunately, fine-tuning on the dinner dataset can enable the model to learn user's preference for dinner more accurately and the user preference learned on the breakfast dataset can assist in the dinner decision as well. In addition, fine-tuning has the following advantages over multi-domain learning: 1) \textbf{model-agnostic}: it can be used to extend any industrial-scale recommender, and 2) \textbf{low training cost}: on the basis of an industrial-scale recommender, fine-tuning the model with one-week-data in the corresponding domain/condition can achieve good performance.

In this paper, we first present a preliminary explanation of fine-tuning in recommender systems. To build a tailored pre-training and fine-tuning (PF) framework for recommendation algorithms, we select our Online Large Recommender (OLR) deployed on an online food platform serving hundreds of millions of users as the pre-trained model. We propose a plug-and-play Information-Aware Adaptive Kernel (IAK) strategy to fine-tune the large pre-trained model. Specifically, the OLR mainly includes five parts: 1) embedding layer \cite{liu2020automated}, 2) short-term and long-term sequence module \cite{manotumruksa2020sequential}, 3) MMoE-like main net (MMoE is a multi-task learning framework \cite{ma2018modeling}), 4) debias module \cite{chen2023bias}, and 5) stacked logits layer. Because the structure of OLR cannot be disclosed in detail due to trade secret, we will also use IAK on the commonly used SOTA baselines to verify the effectiveness of our proposed fine-tuning technique on the recommendation system. IAK is an encoder-decoder structure that can adaptively learn prior knowledge related to downstream tasks with the help of the information bottleneck. IAK compresses general business knowledge at first, then learns specific knowledge on given downstream tasks (In this paper, we consider learning on the given domain/condition as a downstream task.) and forgets the part of business knowledge that is irrelevant to the given task. This approach can draw users more accurately and give proper recommendation results in different domains/conditions flexibly. In addition, we also report two potential issues of fine-tuning in recommendation systems: 1) pseudo cold start issue, 2) user/item overlapping issue, and give preliminary discussions on the concerns.

Our contributions are summarized as follows:
\begin{itemize}
    \item We first present a theoretical explanation of the fine-tuning technique in recommender systems from the viewpoint of the information bottleneck. This makes a significant contribution to interpretability and points out a new topic for future research.
    \item We share unique and important lessons we have learned when deploying the presented solution on a billion-user scale online food platform.
    \item We first report two hidden issues of the fine-tuning technique in recommender systems and provide explorations and discussions.
\end{itemize}

Extensive offline and online experiments on multiple real-world datasets show the superiority of our proposed approaches.

\section{related works}
There are mainly three issues related to this work: multi-domain learning, fine-tuning technique and multi-task learning. 

\textbf{Multi-Domain Learning:} Multi-domain recommendation \cite{zhang2016multi,luo2022mamdr,qian2023adaptive} aims to improve recommendation performance in each domain using knowledge transferred from other domains. SAR-Net \cite{shen2021sar} employs the paradigm of a mixture of experts to give recommendation results and accommodates an attention mechanism \cite{dong2021attention} to extract the relationship of the user's interests in different domains. STAR \cite{sheng2021one} presents a star topology including shared parameters and domain-specific parameters, which can realize using one model to serve all domains in CTR prediction. AFT \cite{hao2021adversarial} learns the feature translations between different domains under a generative adversarial network \cite{goodfellow2014generative} framework. It explicitly models the relationships between items, domains and users' representations in general and specific domains. ADIN \cite{jiang2022adaptive} is able to adaptively treat the commonalities and diversities across scenarios, tracing the user's interest in different domains with shared networks and specific networks. HKGCL \cite{li2023hkgcl} handles each relevant domain as a hierarchy in the interaction network, and the hierarchical knowledge graph is aggregated based on the LightGCN aggregation strategy to learn knowledge representations for users and items.

\textbf{Fine-Tuning in Recommendation:} Although fine-tuning techniques are widely used in NLP, there are few studies on fine-tuning in recommendation. \cite{zhu2022fighting} employs a simple fine-tuning technique to address mainstream bias \cite{li2023mitigating} in recommendation. C$^2$-CRS \cite{zhou2022c2} employs user's feedback to fine-tune a conversational recommender system \cite{sun2018conversational}. 

\textbf{Multi-Task Learning:} Multi-task learning \cite{zhang2018overview, zhang2021survey} is a learning structure that allows for the simultaneous learning of multiple related tasks. Share-Bottom \cite{caruana1997multitask} network designs a shared network at the bottom to learn the similarities and multiple task-specific towers at the top to give the final results. ESMM \cite{ma2018entire} proposes an entire space estimation method for conversion rate prediction and considers the decision path of ``impression->click->conversion". MMoE \cite{ma2018modeling} consists of multiple gates to mix the results of multiple experts and learn the correlations and differences in different tasks. PLE \cite{tang2020progressive} presents a stacked structure with shared experts and specific experts which can decouple the parameters for different tasks to improve the performance.

In contrast to the existing studies, this paper proposes a pre-trained multi-task learning model and utilizes fine-tuning technique to solve multi-domain learning issues. Additionally, this paper provides theoretical exploration and practical guidance through mathematical derivation and extensive experiments.

\section{preliminary}
In this paper, we use the ideology of information bottleneck \cite{tishby2000information, shamir2010learning} to explain the fine-tuning technology. We first introduce information bottleneck in this section.

Information bottleneck theory is originally developed from the rate-distortion theory of data compression in information theory \cite{ash2012information, wu2020graph}, which discusses the problem of how to retain as much information as possible when compressing data. To open the black box of deep neural network, information bottleneck for deep learning is proposed as follows \cite{tishby2015deep, saxe2019information}.

\begin{equation}
    \hat{X} =\text{arg min}~-I(\hat{X};Y)+\beta I(\hat{X};X),
\end{equation}
where $X$ is input data, $Y$ is input label, $\hat{X}$ is minimum sufficient representation of $X$, $\beta$ is a weight, and mutual information $I$ is defined as:
\begin{equation}
    I(X;Y)=\int_Y\int_X p(x,y)\log\left(\frac{p(x,y)}{p(x)p(y)}\right)dxdy.
\label{eq2}
\end{equation}

Further, a variational approximation form with neural network is proposed to make information bottleneck become more suitable for deep learning analysis.
\begin{equation}
\begin{split}
    L=\frac{1}{N}\sum_{i=1}^N\int d\hat{X}p(\hat{X}|X_i)\log q(Y_i|\hat{X}) \\+\beta D_{KL}(p(\hat{X}|X_i)\Vert r(\hat{X})),
\end{split}
\end{equation}
where $N$ is the number of data, $q(Y_i|\hat{X})$ and $r(\hat{X})$ are variational approximation to $p(Y_i|\hat{X})$ and $p(\hat{X})$, and $D_{KL}$ is KL-divergence formulated as:
\begin{equation}
    D_{KL}(p(x)\Vert q(x))=\sum_ip(x_i)\log\left(\frac{p(x_i)}{q(x_i)} \right).
\end{equation}

Information bottleneck for deep learning describes how, during the training process, the neural network retains label information and compresses the data as much as possible to learn the representation from the massive data. For example, the feature map generated by convolutional layers in a convolutional neural network greatly compresses the information of the original data, but retains important features that can identify its labels. The user interest vector extracted by the user historical behavior sequence extraction module in the sequence recommendation algorithm greatly compresses the user behavior information, but retains the main interest of the user.

There are also some studies on the information bottleneck for machine learning \cite{goldfeld2020information, wu2020learnability}. However, the  information bottleneck for fine-tuning technique was not considered by the early research. 

\section{Explanation of Fine-tuning in Recommender}
\subsection{Information Bottleneck for Fine-tuning}
Here, we present a new explanation of fine-tuning from the perspective of information bottleneck. The original information bottleneck describes a minimum sufficient representation of samples that contains specific information. OLR is pre-trained on all domain data on a large scale to fully learn the general business knowledge, but it is difficult to deal with the impact of sudden changes in user interest caused by changes in conditions in downstream tasks (\textit{e.g.} human habits determine that there is a huge gap between user-item distribution of breakfast and 
that of dinner). Thus, each downstream task has its hidden characteristics which is the learning objective of fine-tuning. This means that some of the general business knowledge learned by the pre-trained model is helpful in understanding the downstream task, and some is not. Therefore, the knowledge or information learned in the pre-trained model needs to be compressed, and the information related to the downstream task should be retained as much as possible. 

\textbf{Theorem 1} (Information Bottleneck for Fine-tuning). \textit{Given the general knowledge in the pre-trained model $G$ and the knowledge in the given downstream task $T$, the 
optimization objective of Information Bottleneck for Fine-tuning is to find the minimum sufficient representation $\hat{G}$:}
\begin{equation}
    \hat{G}=\text{arg min}-I(\hat{G};T)+\beta I(\hat{G};G).
    \label{eq6}
\end{equation}

\subsection{Two Phases in Fine-tuning}
It is generally believed that fine-tuning is a process of knowledge transfer \cite{thompson2006clarifying, min2017knowledge}. According to the above discussion, this process can be divided into two parts: knowledge compression and knowledge matching. Knowledge compression describes the process of extracting knowledge from general knowledge that is relevant to the downstream task. Knowledge matching describes the process of supplementing missing knowledge with downstream task data and integrating old and new knowledge.

\textbf{Theorem 2} (Knowledge Compression). \textit{Given the general knowledge in the pre-trained model $G$ and the knowledge in the given downstream task $T$, the optimization objective of Knowledge Compression can be expressed as: }
\begin{equation}
    \Bar{G}=\text{arg min}~D_{KL}(\Bar{G}\Vert T).
\end{equation}

\textbf{Theorem 3} (Knowledge Matching). \textit{Given the compressed general knowledge $\Bar{G}$, the knowledge in the given downstream task $T$, and the learner $f$, the optimization objective of Knowledge Matching can be expressed as follows:}
\begin{equation}
    \hat{G}=\text{arg min}~D_{KL}(\Bar{G}+f(T; \Bar{G})\Vert T).
\end{equation}

\subsection{Approximated Upper Bound}
In light of the above discussion, we calculate the upper bound of the information bottleneck for fine-tuning. In \eqref{eq6}, the first term can be expressed as follows.
\begin{equation}
\begin{split}
    -I(\hat{G};T)=H(\hat{G}|T)-H(\hat{G}),
    \end{split}
\end{equation}
where $H$ means entropy presented as follows. 
\begin{equation}
    H(p(X),q(X))=\int -p(X)\log q(X)dX.
\end{equation}

Suppose the information entropy of $\hat{G}$ is fixed. The objective becomes to minimize the cross-entropy between $\hat{G}$ and $T$. Alternatively, we can optimize the cross-entropy loss of estimation and label directly:
\begin{equation}
    -I(\hat{G};T)\approx H(\hat{Y}|Y),
\end{equation}
where $Y$ is label in $T$ and $\hat{Y}$ is estimation given by fine-tuned model.

For the second term, based on the non-negative nature of KL divergence, the following derivation is valid:
\begin{equation}
\begin{split}
    \int p(\hat{G})\log p(\hat{G})d\hat{G} \geq \int p(\hat{G})\log r(\hat{G})d\hat{G},
\end{split}
\label{eq11}
\end{equation}
where $r(\hat{G})$ is the variational approximation to $p(\hat{G})$.\footnote{It is difficult to calculate $p(\hat{G})$ directly.}

Thus, the upper bound of \eqref{eq6} is as follows.
\begin{equation}
\begin{split}
    -I(\hat{G};T)+\beta I(\hat{G};G) 
    \leq H(\hat{Y}|Y)+\frac{\beta}{N}\sum_{i=1}^{N}\left[\log\left(\frac{p(\hat{G}_i|G_i)}{r(\hat{G})}\right) \right],
\end{split}
\end{equation}
where knowledge $G$ and $\hat{G}$ can be approximated with training samples. More details are presented in Appendix A.1.

\section{proposed method}
\subsection{Problem Formulation}
In this paper, we mainly focus on multi-scene, multi-region and multi-period in food recommendations. Let $S, R, P$ denote three topics, respectively. Given a set of downstream tasks $D=S \cup R \cup P=\{s_1,...,s_i\}\cup \{r_1,...,r_j\}\cup\{p1,...,p_k\}$ where $s_i$ denotes the i-th scene, $r_j$ denotes the j-th city, and $p_k$ denotes the k-th mealtime. Here, a task is defined as modeling user interest in the given domain/condition. Therefore, there are $i+j+k$ downstream tasks, the tasks under the same topic are independent, and the tasks under different topics are orthogonal. $G$ is the general knowledge and $T_m$ is specific knowledge in the m-th task. The challenge is how to effectively extract the knowledge related to $T_m$ from $G$ to assist the model in decision-making on the m-th task, in addition, to accomplish the above goal while compressing business knowledge of the large model as much as possible. This can be formulated as follows.
\begin{equation}
    \text{arg min}~\Vert\mathbb{F}_\theta(G)-T_m\Vert+|G|,
\end{equation}
where $\mathbb{F}_\theta$ is the learner with parameter $\theta$ and $|G|$ expresses the information of business knowledge. 

\subsection{Online Large Recommender}
Here, we give a brief introduction to the OLR, which serves billions of recommendation requests per day.\footnote{The model involves trade secrets and cannot be disclosed in detail.} OLR is a multi-task schema flow trained on the comprehensive domain dataset, allowing OLR to extract general business knowledge and track user and item profiles. First, all ID features and discretized continuous features are input into an embedding layer to generate dense low-dimensional vectors. Once the data is embedded, the historical behavior related features are fed into short-term and long-term sequence modules to draw the interest evolution and user-item interaction, ensuring personalized mining. Then, the extracted user interest vectors are merged with other features and fed into an MMoE-like main net to learn the underlying relationship between CTR and CTCVR. In addition, a debias module is employed to avoid position bias, user bias, \textit{.etc}, to assist OLR in providing a ground-truth result. Finally, the stacked logits layer uses all features and knowledge to perform estimation. The number of parameters of OLR is around one billion and it is pre-trained on hundreds of billions of samples.

\subsection{Information-Aware Adaptive Kernel}
Although we have provided an upper bound for the information bottleneck for fine-tuning, it is not easy to apply it to model optimization. Thus, we further relaxed the upper bound to simplify it.
\begin{equation}
\begin{split}
    \hat{G} \leq H(\hat{Y}|Y)-\beta\log r(\hat{G}).
\end{split}
\end{equation}

Since we have to minimize $\hat{G}$, minimizing its upper bound is a good alternative. More details are presented in Appendix A.2.
\begin{equation}
    \hat{G}=\text{arg min}~H(\hat{Y}|Y)-\beta\log r(\hat{G}).
\end{equation}

IAK is designed as an encoder-decoder structure. In this paper, we employ MLPs as the encoder and decoder. The encoder is mainly responsible for knowledge compression and both the encoder and decoder are responsible for knowledge matching. We consider the output of the encoder to represent the compressed business knowledge of the large model. 

\textbf{Gaussian Approximation} It is still difficult to solve $-\beta\log r(\hat{G})$ directly, so we propose an alternative. Starting from the original meaning of the information bottleneck, the second term's requirement is to compress the original information as much as possible. We can address this problem from the distribution of parameters. If the parameter distribution of the well-trained model is similar to the initial parameter distribution, it indicates that the input sample has little influence on the model, that is, the knowledge compression rate is large. Here we employ a standard Gaussian distribution to initialize the encoder's parameters $w_0$: 
\begin{equation}
    w_0\sim \mathcal{N}(0,\mathbf{I}).
\end{equation}

In order to optimize the second term, we consider a parametric Gaussian approximation:
\begin{equation}
    w\sim \mathcal{N}(\mu,\sigma),
\end{equation}
where $\mu$ and $\sigma$ are mean and variance of parameters of the encoder, and $w$ is the well-trained parameters. 

The distance $\mathcal{D}$ between the distribution of the trained parameter and that of the original parameter can be formulated as the following equation.
\begin{equation}
    \mathcal{D}=D_{KL}(w\Vert w_0).
\end{equation}

Based on the above assumptions, the objective can be simplified as follows. More details are given in Appendix B.
\begin{equation}
    \text{arg min}~H(\hat{Y}|Y)+D_{KL}(w\Vert w_0).
\end{equation}
Here, we make the following assumption.

\textbf{Assumption 1}. \textit{The distance between the distribution of the trained parameter and that of the original parameter is positively correlated with the information contained in the parameter.} 

During the training stage, the parameters of the large pre-trained model are frozen to maintain the general business knowledge and we only train IAKs on the downstream datasets.

\subsection{Lessons Learned from the real recommender}
In a real recommendation system, deploying a solution on the server presents many limitations and challenges. Here, we share two important lessons learned in our real business to help practitioners use IAK technique in their recommenders.

\subsubsection{Parallel Inference and Domain Activation} For online inference, each model needs to be deployed independently and consumes server resources. Therefore, we cannot deploy the corresponding IAK on each domain. We propose a parallel inference and domain activation strategy to address this problem. All IAKs for different domains are deployed simultaneously. For any recommendation request, all IAKs will infer in parallel, and then select the output of the corresponding IAK according to the domain ID, which can be formulated as:
\begin{equation}
\mathbb{F}=\sum_i^{N_d}\mathbb{F}_i\times\mathbb{I}(d=d_i),
\end{equation}
where $\mathbb{F}$ represents the IAK, $N_d$ is the number of domains, $\mathbb{F}_i$ represents the IAK for the i-th domain, $d$ is the domain ID of the current recommendation request, and $\mathbb{I}$ is the indicator function.

\subsubsection{Dynamic Batch Aware Training}
Although IAK is easy to use, training multiple IAKs for different downstream tasks is cumbersome. Moreover, due to the varying number of samples in different downstream tasks, it is difficult to combine multiple downstream tasks for training. This may lead to a lack of data for some downstream tasks in a training batch, resulting in large fluctuations in the corresponding IAK's gradient, which affects convergence. Therefore, the learning rate should change with the number of samples and the magnitude of the gradient.
\begin{equation}
    \hat{\lambda} = \text{SoftMax}(N_B\odot W)\times \lambda,
\end{equation}
where $\lambda$ is the original learning rate, $N_B$ is the number of samples in a batch for IAKs, $W$ is the reciprocal of the gradient for IAKs and $\odot$ denotes element-wise product. We still recommend choosing tasks of a similar magnitude to train together, or training them one-by-one.

\section{discussion on the hidden issues}
In this section, we report two concerns about the fine-tuning technique in recommender systems. These issues are significant for understanding the role of fine-tuning in the field of recommendation systems.

\textbf{Pseudo cold start issue.} In practice, the amount of data used for fine-tuning is much smaller than that used for pre-training, and obviously the historical records of user-item interactions in some domains are lost in the downstream dataset. For example, we select the last seven days of scene A data to fine-tune OLR, but user B has not entered scene A in the last seven days. The IAK module does not learn the user's behavior in scene A from the downstream dataset. When the model is deployed on the server, the user B enters scene A, which the IAK module encounters for the first time, creating a pseudo cold start issue\footnote{Unlike the traditional cold start problem \cite{lika2014facing}, this user is not an absolutely new user, but a new user for a certain period of time in a specific scene.}. Models can only estimate based on the pre-trained general knowledge.

\textbf{User/item overlapping issue.} Traditional fine-tuning methods completely isolate samples for different domains. Models serving scene A are fine-tuned using only scene A's samples. But that doesn't make sense in the realm of recommendation systems. In a recommendation system, samples from different domains are highly correlated. Taking multi-period as an example, there is not much difference in the distribution of items for lunch and dinner. So the information of an item at noon is helpful for understanding the characteristics of the item at night. It's not appropriate to think of them as independent tasks.

\section{experiments}
We conducted extensive experiments to evaluate the performance of IAK, and the following research questions (RQs) were answered:

\begin{itemize}
    \item \textbf{RQ1} Does OLR outperform state-of-the-art baselines in multiple conditions?
    \item \textbf{RQ2} Does IAK fine-tuning technique further improve the performance of OLR?
    \item \textbf{RQ3} Does IAK fine-tuning technique further improve the performance of baselines?
    \item \textbf{RQ4} Can IAK address multiple topics at the same time?
    \item \textbf{RQ5} How do critical components affect the performance of IAK?
    \item \textbf{RQ6} How does the issue in Section 6 affect IAK's performance?
    \item \textbf{RQ7} Does OLR+IAK work in real large-scale online recommendation scenarios?
    \item \textbf{RQ8} How does IAK affect recommendation results?
\end{itemize} 

\subsection{Experimental Settings}
Here we introduce our experimental settings. More details are given in Appendix C. 
\subsubsection{Datasets}
We collected 11 offline datasets from the user's behavior logger between July 20, 2023 and July 27, 2023 on a large-scale online platform serving over one billion recommendation requests per day. The datasets can be classified into three groups according to topics: 1) multi-region, 2) multi-period, and 3) multi-scene. The statistics of datasets are summarized in Table~\ref{tab1}.

\begin{table*}[t]
    \centering
    \caption{The basic statistics of datasets. ``\#" denotes ``the number of". K refers to thousand and M refers to million.}
    \begin{tabular}{ccccccccc}
    \toprule
      Group & Dataset & \#Users & \#Items & \#Clicks & \#Purchases & Total Size & Sparsity of Click & Sparsity of Purchase \\
    \hline
      Multi-scene & Scene1& 63M & 3M & 273M & 38M & 4,811M & 5.69\% & 0.79\%\\
      & Scene2 & 22M & 2M & 111M & 15M & 1,738M & 6.43\% & 0.89\%\\
      \hline
      & Time1 & 16M & 2M & 31M & 5M & 533M & 5.99\% & 0.91\%\\
      Multi-period & Time2 & 38M & 3M & 133M & 20M & 2,199M & 6.04\% & 0.92\%\\
      & Time3 & 38M & 3M & 128M & 17M & 2,158M & 5.92\% & 0.79\%\\
      \hline
      & Region1 &5M  & 120K & 42M & 6M & 718M & 5.95\% & 0.91\%\\
      & Region2 &2M & 60K & 17M & 2M & 284M & 5.99\% & 0.92\% \\
      Multi-region & Region3 &3M & 91K & 20M & 3M & 343M & 5.96\% & 0.87\%\\
      & Region4 &2M & 73K & 8M & 1M & 142M & 5.86\% & 0.75\%\\
      & Region5 &1M & 33K & 8M & 1M & 146M & 5.74\% & 0.81\%\\
      & Region6 &1M & 37K & 8M & 1M & 129M & 6.12\% & 0.87\%\\
    \bottomrule
    \end{tabular}
    \label{tab1}
\end{table*}

\subsubsection{Baselines}
In most large-scale practices, the core goals of recommenders are the estimation of click-through rate (CTR) and click-through conversion rate (CTCVR) \cite{wen2020entire, jiang2023esmc}. Therefore, the discussion in this paper is based on the multi-task learning framework. The representative state-of-the-art approaches are listed as follows.
\textbf{Shared Bottom (SB)} \cite{caruana1997multitask} is one of the most classic multi-task models. The bottom learns the underlying relationships among different tasks and each task corresponds to an output tower. \textbf{ESMM} \cite{ma2018entire} estimates CTR and CTCVR via explicitly modeling the user's decision path of "impression-click-conversion" with the chain rule of conditional probability. \textbf{MMoE} \cite{ma2018modeling} employs multiple experts and gates to leverage different tasks. It uses gates to adapt the weights of experts on different tasks.

\subsubsection{Metrics}
To evaluate the performance of the proposed approach, we select two widely used metrics for offline testing, \textit{i.e.}, CTR-AUC and CTCVR-AUC for CTR and CTCVR estimation, respectively.

\begin{table}[t]
    \centering
    \caption{Comparison with SOTA baselines in terms of CTR-AUC. ZS-OLR means zero shot OLR.}
    \begin{tabular}{cccccc}
    \toprule
       Dataset & SB & ESMM & MMoE & ZS-OLR & OLR+IAK\\
    \hline
       Scene1&0.70034 &0.70452 &0.71133 &\textit{0.71405} &\textbf{0.72192}\\
       Scene2 &0.72067 &0.71986 &0.72131 &\textbf{0.73045} &\textit{0.72635}  \\
        Avg. &0.71051 &0.71219 &0.71632 & \textit{0.72225} & \textbf{0.72414} \\
      \hline
       Time1 &0.73246 &0.73525 &0.74103 &\textit{0.74597}&\textbf{0.74941} \\
      Time2 &0.70008 &0.70852 &0.70996 &\textit{0.71263} &\textbf{0.72004} \\
       Time3 &0.70543 &0.71169 &0.71471 &\textit{0.72042} &\textbf{0.72392}  \\
       Avg. &0.71266 &0.71849 &0.72190 & \textit{0.72634}& \textbf{0.73112} \\
      \hline
       Region1 &0.72093 &0.72322 &0.72614 &\textit{0.72978} &\textbf{0.73129} \\
       Region2 &0.72081 &0.72308 &0.72314 &\textit{0.72374} &\textbf{0.72382} \\
       Region3 &0.72998 &0.73013 &0.73106 &\textit{0.73129} &\textbf{0.73180} \\
       Region4 &0.70102 &0.70324 &0.70381 &\textit{0.70784} &\textbf{0.72111} \\
       Region5 &0.71495 &0.71643 &0.72018 &\textit{0.72559} &\textbf{0.72561} \\
       Region6 &0.72945 &0.72957 &0.73004 &\textit{0.73012} &\textbf{0.73016} \\
       Avg. &0.71952 &0.72094 &0.72239 &\textit{0.72473} & \textbf{0.72730}\\
    \bottomrule
    \end{tabular}
    \label{tab2}
\end{table}

\begin{table}[t]
    \centering
    \caption{Comparison with SOTA baselines in terms of CTCVR-AUC. ZS-OLR means zero shot OLR.}
    \begin{tabular}{cccccc}
    \toprule
       Dataset & SB & ESMM & MMoE & ZS-OLR & OLR+IAK\\
    \hline
       Scene1&0.83128 &0.83297 &0.83732 &\textit{0.84084} &\textbf{0.84477}\\
       Scene2 &0.83039 &0.83111 &0.83204 &\textit{0.83210} &\textbf{0.83629}\\
       Avg. &0.83084 &0.83204 &0.83468 &\textit{0.83647} &\textbf{0.84053} \\
      \hline
       Time1 &0.86312 &0.86910 &0.86917 &\textit{0.87394}&\textbf{0.87426} \\
      Time2 &0.81679 &0.82043 &0.82109 &\textit{0.82494} &\textbf{0.83101} \\
       Time3 &0.83011 &0.83327 &0.83509 &\textit{0.83860} &\textbf{0.83892}  \\
       Avg. &0.83667 &0.84093 &0.84178 &\textit{0.84583}&\textbf{0.84806}\\
      \hline
       Region1 &0.83076&0.83548&0.83602&\textit{0.83947} &\textbf{0.83957} \\
       Region2 &0.82431 &0.82887 &0.82896 &\textit{0.83021} &\textbf{0.83031} \\
       Region3 &0.83951 &0.84202 &0.84309 &\textit{0.84347} &\textbf{0.84362} \\
       Region4 &0.81960 &0.82338 &0.82410 &\textit{0.82744} &\textbf{0.83233} \\
       Region5 &0.83072 &0.83475 &0.83512 &\textit{0.83891} &\textbf{0.83894} \\
       Region6 &0.83790 &0.84202 &0.84441 &\textit{0.84618} &\textbf{0.84635} \\
       Avg.& 0.83047&0.83442&0.83528&\textit{0.83761} &\textbf{0.83852} \\
    \bottomrule
    \end{tabular}
    \label{tab3}
\end{table}

\subsection{RQ1\&RQ2: Comparison with Baselines}
In this subsection, we discuss the results of baselines \textit{vs.} our models and OLR \textit{vs.} OLR+IAK to answer RQ1 and RQ2.

Table~\ref{tab2} and Table~\ref{tab3} indicate that OLR outperforms all state-of-the-art (SOTA) baselines in terms of CTR-AUC and CTCVR-AUC on 11 real-world datasets. The best results are shown in \textbf{Bold} and the second best results are shown in \textit{Italic}. All the best results and the second best results come from our methods. Compared to the baselines, OLR+IAK achieves a huge improvement on average, \textit{i.e.}, CTR-AUC +1.09\% and CTCVR-AUC +0.70\% for multi-scene, CTR-AUC +1.28\% and CTCVR-AUC +0.75\% for multi-period, and CTR-AUC +0.68\% and CTCVR-AUC +0.39\% for multi-region. The improvement verifies the superiority of our proposed approaches.

Considering ZS-OLR and baselines, MMoE achieves the best performance among the baselines. However, our OLR still makes a great improvement on the basis of MMoE. This shows that OLR has fully learned the general business knowledge and can give accurate recommendations in various downstream tasks without fine-tuning.

For ZS-OLR and OLR+IAK, OLR+IAK outperforms ZS-OLR on all datasets except comparison on Scene2 in terms of CTR-AUC. Compared to ZS-OLR, OLR+IAK achieves a significant improvement on average, \textit{i.e.}, CTR-AUC +0.26\% and CTCVR-AUC +0.49\% for multi-scene, CTR-AUC +0.66\% and CTCVR-AUC +0.26\% for multi-period, and CTR-AUC +0.35\% and CTCVR-AUC +0.11\% for multi-region. This indicates that IAK can further improve the recommendation performance of OLR on multiple domains. Such a big improvement on a high benchmark is astounding and proves that IAK is meaningful and promising.

\subsection{RQ3: Control Test}
To demonstrate the universality of IAK, we use it to fine-tune the baseline models. 

\begin{figure}[t]
    \centering
    \subfloat[Results of control test in terms of CTR-AUC.]{\includegraphics[width=0.94\columnwidth]{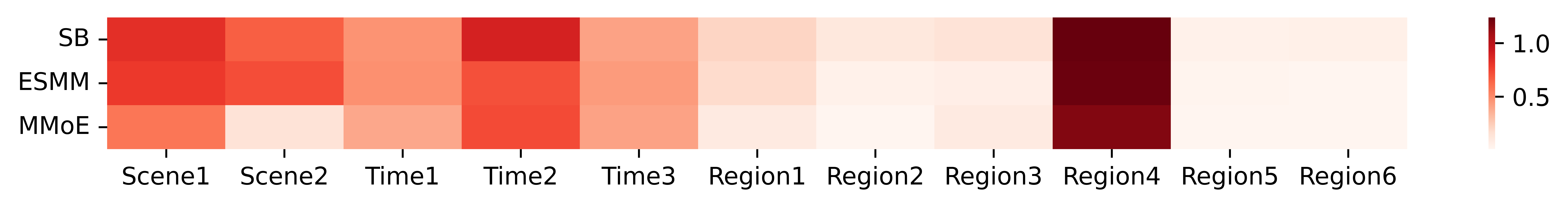}}\\
    \subfloat[Results of control test in terms of CTCVR-AUC.]{\includegraphics[width=0.96\columnwidth]{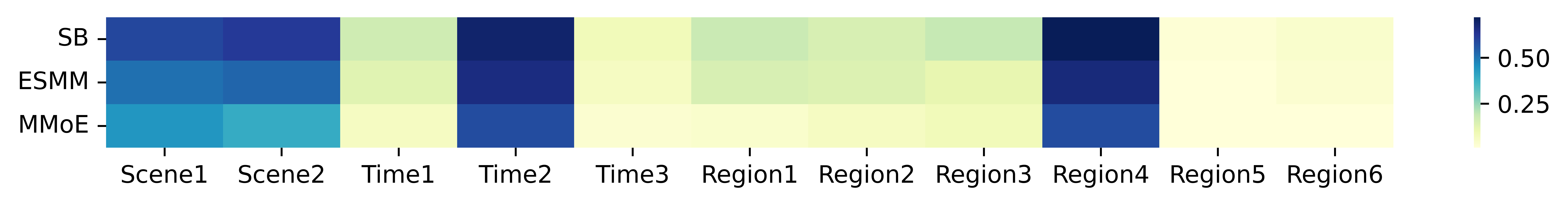}}
    \caption{Heatmap of results of control test.}
    \label{fig-control}
\end{figure}

Fig.~\ref{fig-control} visualize the results of the control test. We can see that IAK fine-tuning technique consistently improves the performance of the baseline models. This proves that IAK is a universal module that can be employed in any model. Additionally, IAK performs better in Scene1, Scene2, Time2, and Region4 which is consistent with the results of fine-tuning OLR. In light of the results of data analysis, there is a large gap between the data distribution of the better performing domains and the overall data distribution. In these domains, the user's interest shifts severely and discontinuously, making it difficult for the model to capture the user's intent. Fine-tuning alleviates this problem by allowing the model to learn domain-specific sample distributions that can explicitly trace the sudden change of the user's interest. As a result, models can perform better on these domains.

\subsection{RQ4: Cross-Topic Test}
In this subsection, we verify the performance of IAK in cross-topic conditions. Here, we combine domains under different topics in pairs to form new domains. Table~\ref{tab-cross} presents the results of cross-topic tests. On average, OLR+IAK achieves CTR-AUC +0.37\%, CTCVR-AUC +0.42\% in Scene-Region, CTR-AUC +0.55\%, CTCVR-AUC +0.61\% in Scene-Time, and CTR-AUC +0.63\%, CTCVR-AUC +0.31\% in Region-Time. The results prove the superiority of IAK.

Cross-topic testing is meaningful for the real recommendation system. In reality, conditions under different topics occur at the same time, and the ability of cross-topic modeling can depict users' interests more accurately and further improve the recommendation performance.

\begin{table}[t]
    \centering
    \caption{Cross-topic test. The results are the relative increase of OLR+IAK compared to ZS-OLR.}
    \begin{tabular}{c|cccc}
    \toprule
    &&CTR-AUC&\\
    \hline
      Domain & Scene1 & Scene2 & Region1 & Region3  \\
    \hline
    Scene1 & / &  / & +0.69\% & +0.72\% \\
    Scene2 & / & / & +0.01\% & +0.05\%\\
    Time2  & +0.94\% & +0.12\% & +0.63\% & +0.64\%\\
    Time3  & +0.98\% & +0.14\% & +0.60\% & +0.65\% \\
    \hline
    &&CTCVR-AUC&\\
    \hline
      Domain & Scene1 & Scene2 & Region1 & Region3  \\
    \hline
    Scene1 & / &  / & +0.43\% & +0.36\%\\
    Scene2 & / & / & +0.47\% & +0.41\%\\
    Time2  & +0.60\% & +0.63\% & +0.36\% & +0.30\%\\
    Time3  & +0.59\% & +0.61\% & +0.33\% & +0.25\%\\
    \bottomrule
    \end{tabular}
    \label{tab-cross}
\end{table}

\subsection{RQ5: Parameter Sensitivity Test}
\begin{figure}[t]
    \centering
    \subfloat{\includegraphics[width=0.45\columnwidth]{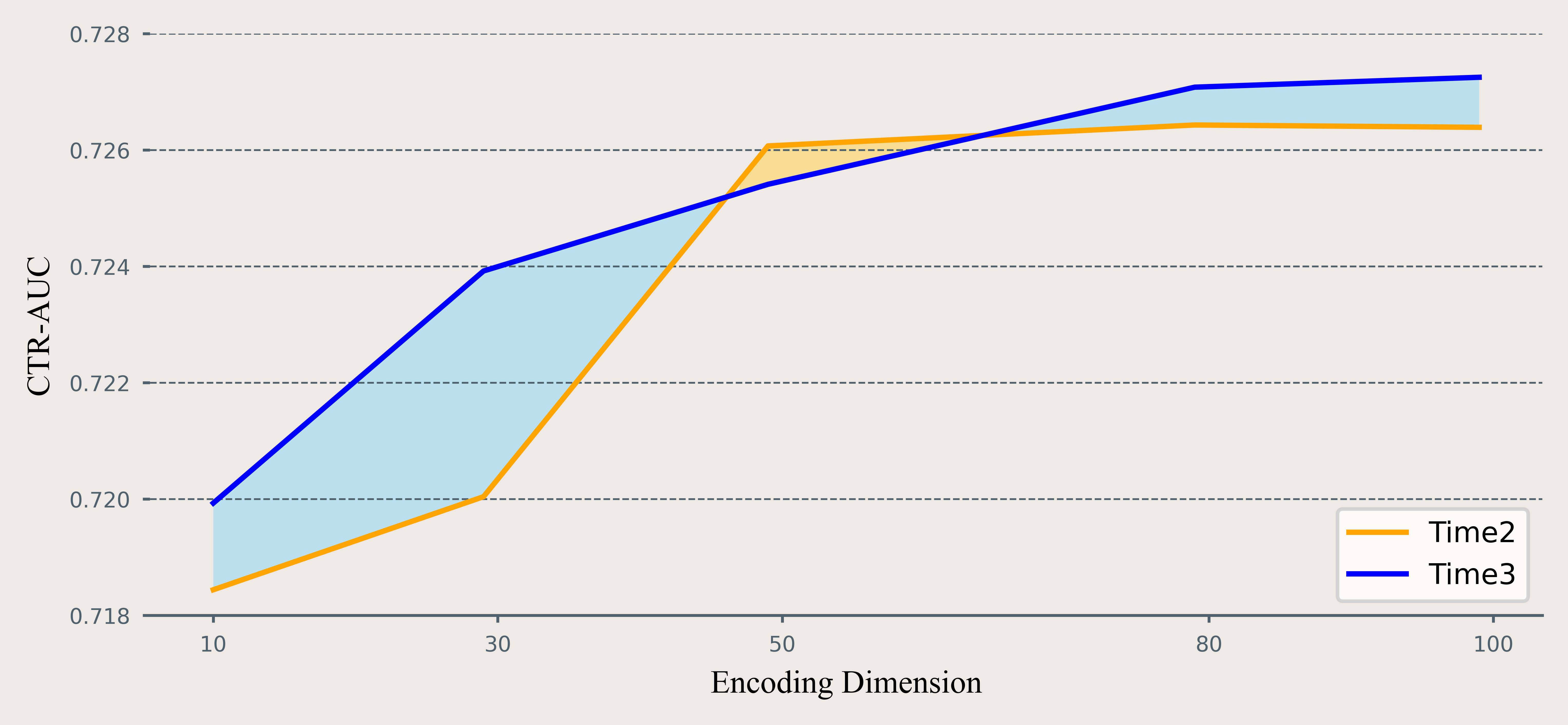}\label{l-a}}\hspace{5pt}
    \subfloat{\includegraphics[width=0.45\columnwidth]{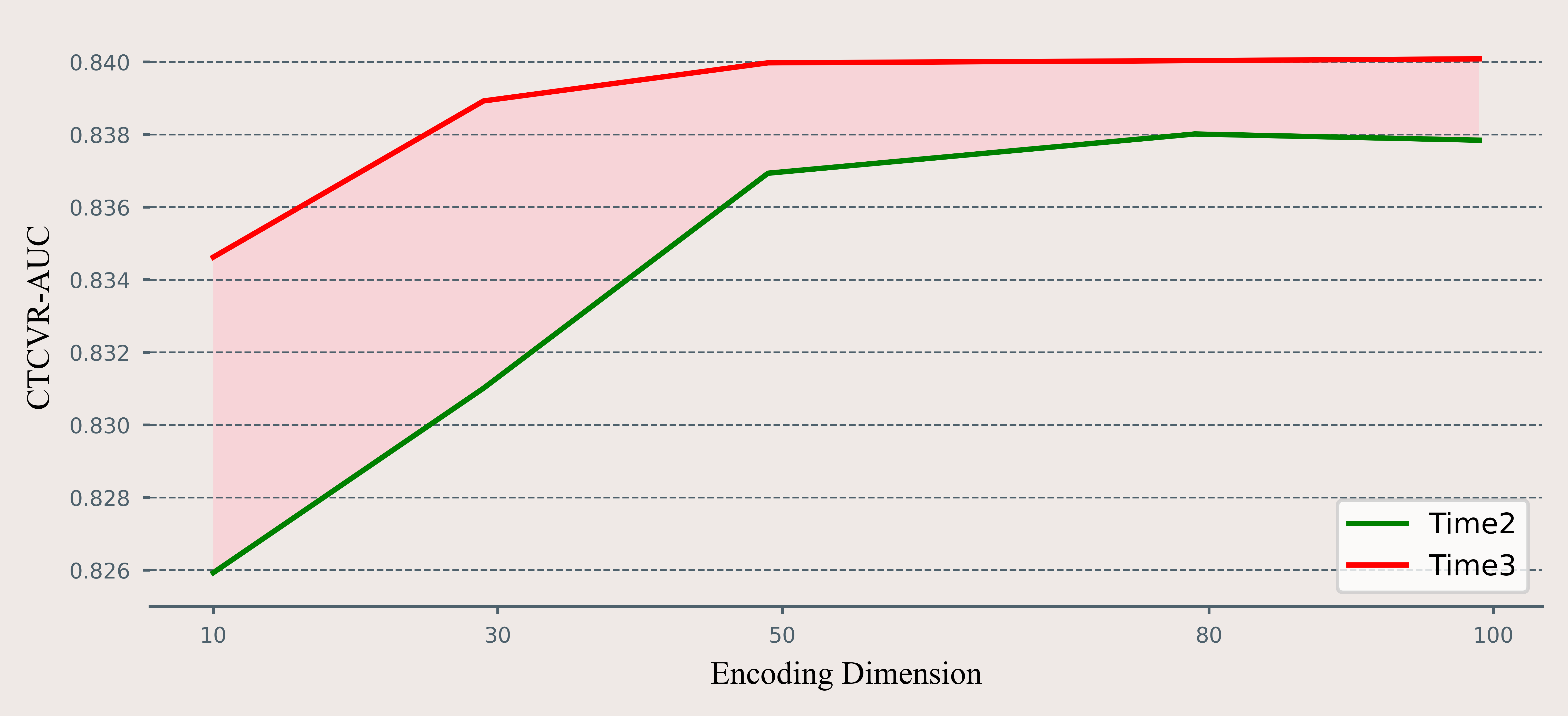}\label{l-b}}
    \caption{Results of parameter sensitivity test.}
    \label{param}
\end{figure}

Since IAK is an encoder-decoder structure, the encoding dimension $d_e$ is an important parameter that affects the IAK structure. Fig.~\ref{param} shows that the performance of the model increases with the increase of $d_e$, while the growth speed is slower and slower. This indicates that there is an upper bound on the number of parameters required to complete the knowledge transfer in IAK. Based on the experiments, we recommend setting $d_e$ to 50.

\subsection{RQ6: Exploration Study}
In this subsection, we implement experiments to preliminarily explore the answers to the concerns proposed in Section 6. More details are presented in Appendix D.

\begin{figure}[t]
    \centering
    \subfloat[Results of exploration of the pseudo cold start issue in terms of CTR-AUC.]{\includegraphics[width=0.45\columnwidth]{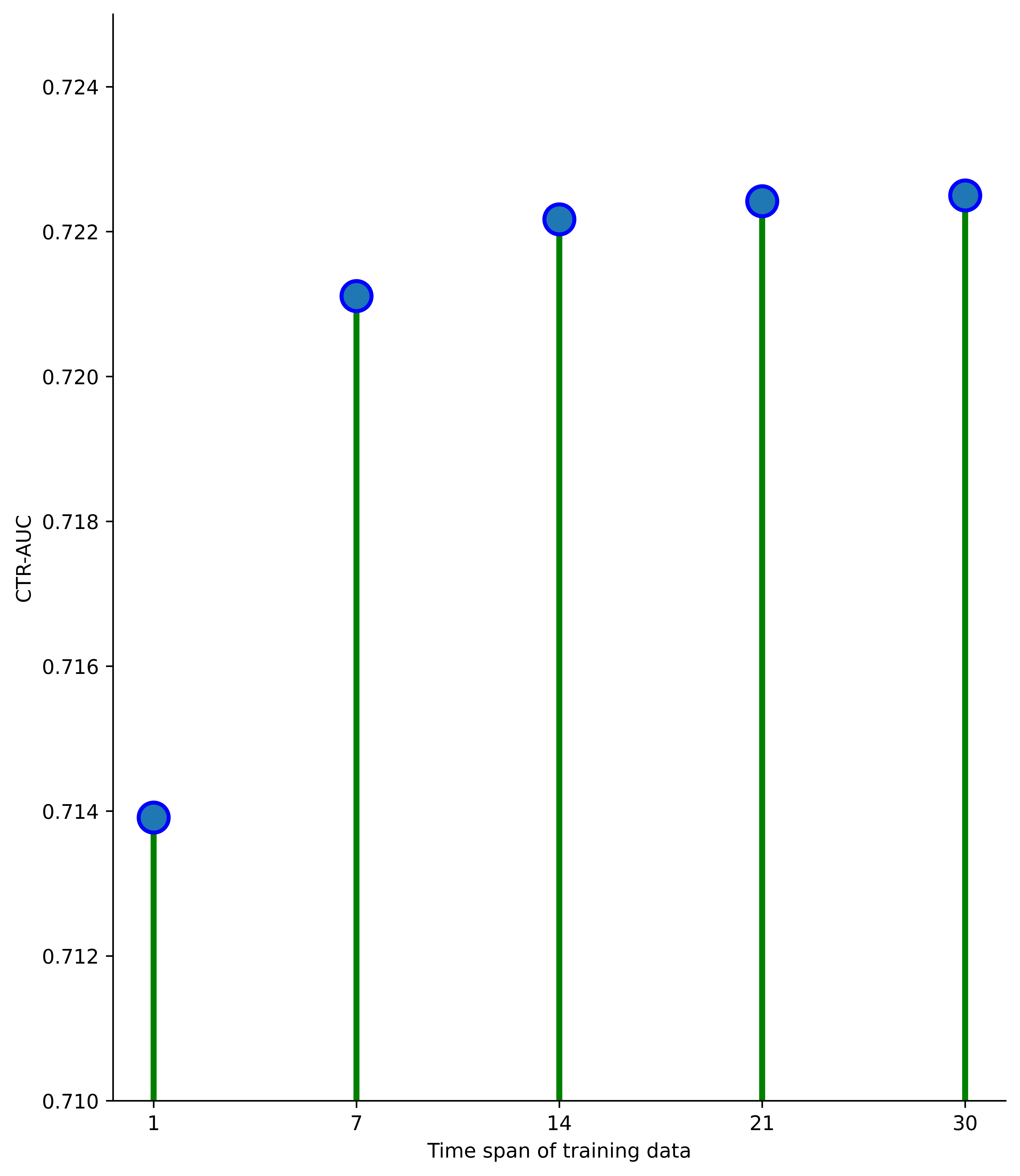}\label{e-a}}\hspace{5pt}
    \subfloat[Results of exploration of the pseudo cold start issue in terms of CTCVR-AUC.]{\includegraphics[width=0.45\columnwidth]{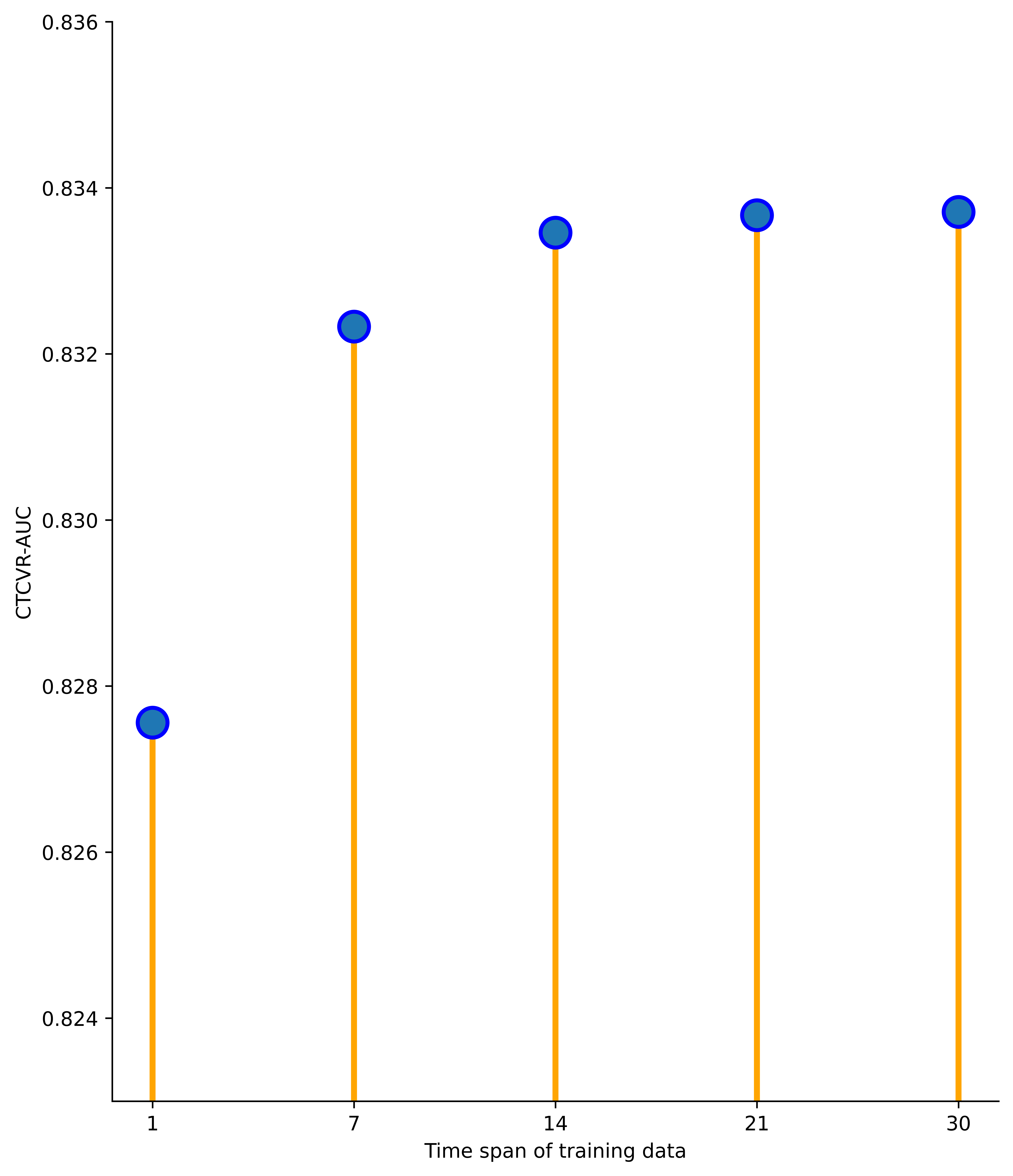}\label{e-b}}\\
    \subfloat[Results of exploration of user/item overlapping issue.]{\includegraphics[width=0.95\columnwidth]{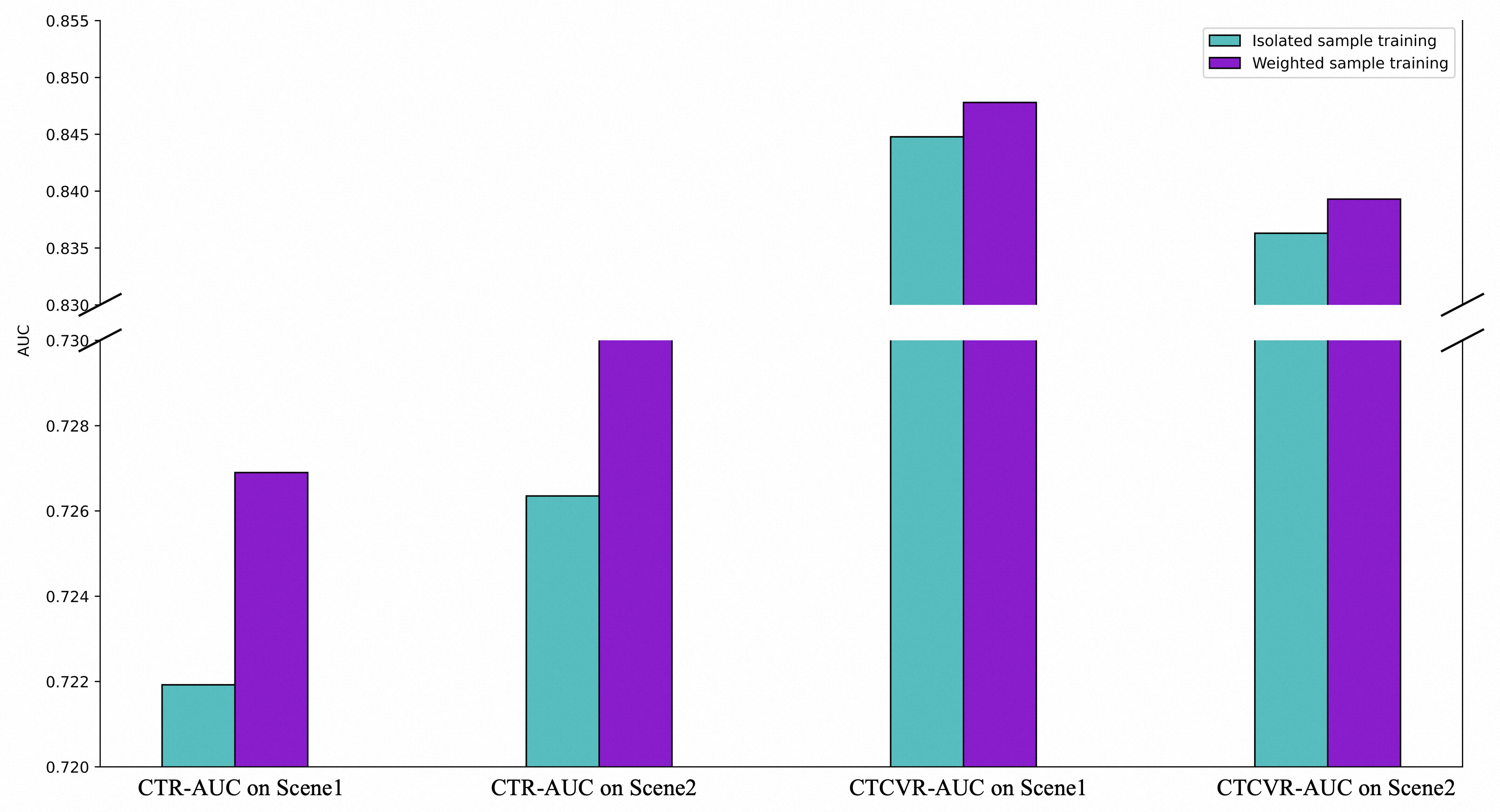}\label{e-c}}
    \caption{Results of exploration study.}
    \label{fig-explore}
\end{figure}

To evaluate the pseudo cold start issue, we test the performance of IAK trained on samples in different time spans. In Fig.~\ref{e-a} and \ref{e-b}, as the time span increases, the performance of the model grows more and more slowly. Even though an increase in time span can further improve model performance, it is not reasonable to increase the time span all the time. Besides, training data over a week-long span is recommended for fine-tuning.

To evaluate the user/item overlapping issue, we try not to completely isolate samples from different domains. For Scene1, we use Scene1 samples with 70\% weight and Scene2 samples with 30\% weight to train IAK for Scene1. The weight setting is based on the proportion of users in real business. Fig.~\ref{e-c} shows that fine tuning on completely isolated samples can not achieve the best performance, and the proposed weighted sample training can further improve the performance of the model by introducing samples from other related domains.

\subsection{RQ7: Online A/B Test}
To evaluate the performance of OLR+IAK in large-scale online recommender, we conducted three online A/B tests (one single IAK test and two multiple IAKs tests) by deploying it to the recommendation scenario on the homepage of a billion-scale online platform. The online base model is ZS-OLR. Here, we selected three business-related metrics: Number of Orders (NO), Order Rate (OR) and Net GMV (NG, a measure of net profit) to evaluate the performance in an online environment. 

\subsubsection{Single IAK Test} 
From August 2, 2023 to August 25, 2023, we conducted a 23-day online A/B test in Region4 serving 2\% of users. On average, OLR+IAK improved NO by 1.86\% OR by 1.14\% and NG by 1.62\% in Region4. This proves the effectiveness of IAK technique in large-scale online recommenders in a given domain. 

\subsubsection{Multiple IAKs Test} To verify the influence of IAK technique on the global domain of recommendation, we conducted experiments in two dimensions: multi-scene and multi-region.

From July 25, 2023 to August 25, 2023, we conducted a 30-day online A/B test in both Scene1 and Scene2 serving 8\% of users. On average, OLR+IAK improved NO by 0.61\% OR by 0.46\% and NG by 0.74\% in the global domain. From August 18, 2023 to August 25, 2023, we conducted a seven-day online A/B test in all regions serving 2\% of users. On average, OLR+IAK improved NO by 0.58\% OR by 0.62\% and NG by 0.74\% in the global domain. The experiments show the superiority of IAK technique in large-scale online recommenders. Finally, we chose to deploy the multi-scene dimensional multi-IAK OLR to the homepage of our platform serving billions of recommendation requests per day.

\subsection{RQ8: Case Study}
In this subsection, we present a representative case to visually illustrate the importance of fine-tuning. 

\begin{figure}[t]
    \centering
    \includegraphics[width=0.95\columnwidth]{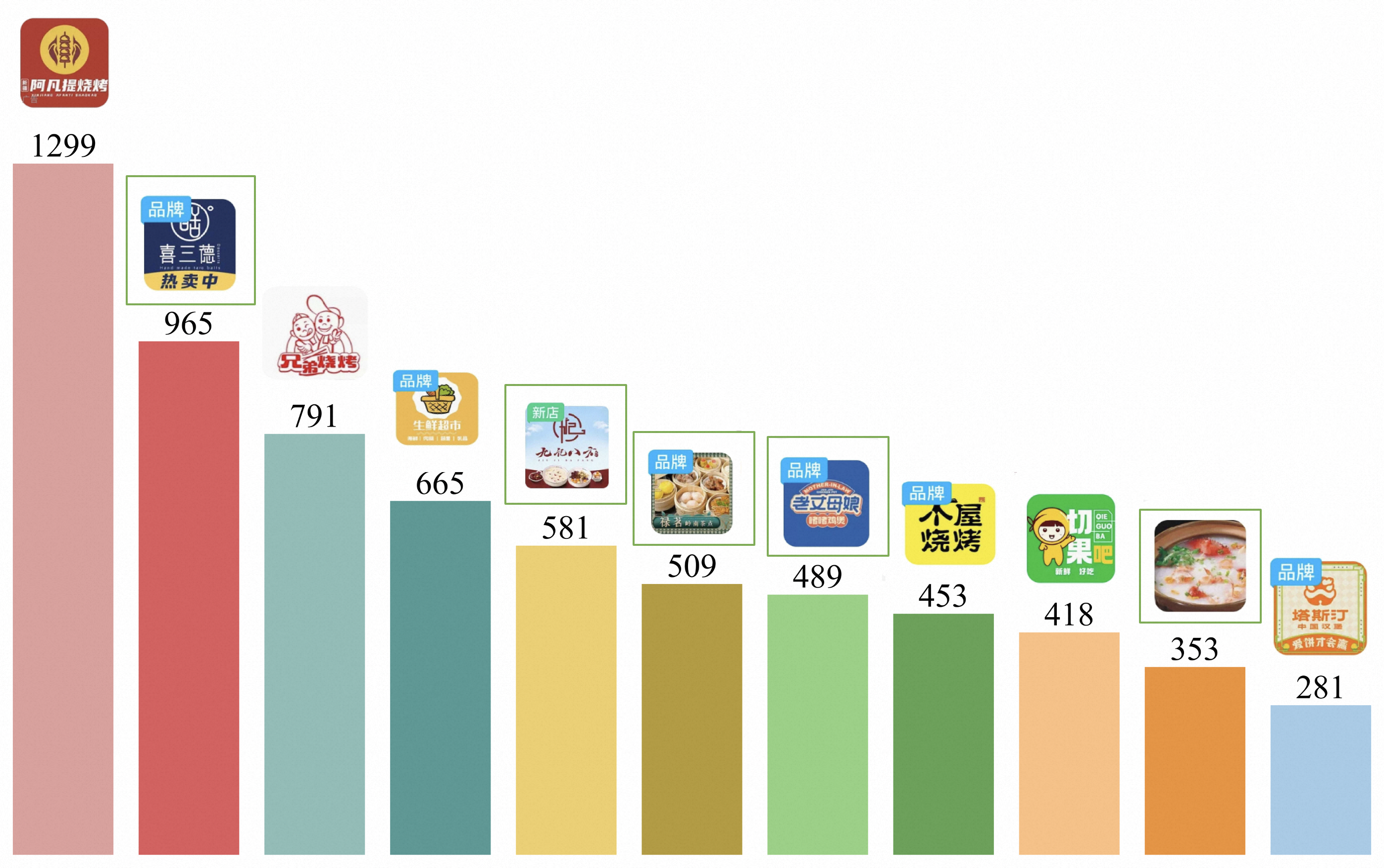}
    \caption{Some hot items in Region4, the vertical axis represents the number of impressions of the items, and the items surrounded by green squares are the specialties of Region4.}
    \label{case}
\end{figure}

Fig.~\ref{case} demonstrates the number of impressions of 11 hot items in Region4. The data is collected from the real online environment deploying OLR+IAK serving 2\% of users in Region4 in seven days. Among the items, five items enclosed by green squares belong to Cantonese cuisine, a specialty of Region4. Users in Region4 are divided into two classes: permanent residents and outsiders. Due to eating habits and living habits, local residents usually prefer local specialties. Outsiders, especially travelers, also prefer local specialties that are regarded as a part of their trip. Besides, barbecue also accounts for a large proportion, which is a popular food in the world. This case demonstrates that IAK technique can mine the characteristics of different domains to make the recommendation results more in line with the user's interest distribution in the downstream task. We can understand the user's interest much better in different conditions in this manner.

\section{conclusion}
In this paper, we propose a fine-tuning technique called IAK for recommenders to address the multi-domain learning issue. To make it suitable for real large-scale recommendation systems, we specifically design deployment and training strategies. Additionally, we first present a theoretical explanation for fine-tuning in recommendation that significantly improves the interpretability of our proposed methodologies. IAK can be flexibly applied to any recommendation model structure without affecting the recommendation performance and overall framework of the original model. Extensive online and offline experiments test the superiority of our proposed approaches. The IAK technique is deployed on the homepage of a real online platform and yields significant financial benefits. Future work may include exploring a better method to mix different domains in different topics.

\begin{acks}
Thanks to Alibaba Group for supporting this study.
\end{acks}

\bibliographystyle{ACM-Reference-Format}
\bibliography{ref.bib}

\appendix
\section{Mathematical Derivation}
\subsection{Mathematical Derivation in Section 4.3}
For the second term, based on the non-negative nature of KL divergence, the following derivation is valid:
\begin{equation}
\begin{split}
    D_{KL}(p(\hat{G})\Vert r(\hat{G}))\geq 0 \\
    \Rightarrow \int (p(\hat{G})\log p(\hat{G})- p(\hat{G})\log r(\hat{G}))d\hat{G} \geq 0 \\
    \Rightarrow \int p(\hat{G})\log p(\hat{G})d\hat{G} \geq \int p(\hat{G})\log r(\hat{G})d\hat{G},
\end{split}
\end{equation}
where $r(\hat{G})$ is the variational approximation to $p(\hat{G})$.

According to \eqref{eq2} and \eqref{eq11}, 
\begin{equation}
\begin{split}
    I(\hat{G};G)&=\int_G \int_{\hat{G}} p(\hat{G},G)\log \left(\frac{p(\hat{G},G)}{p(\hat{G})p(G)}\right)d\hat{G}dG\\
    &=\int_G \int_{\hat{G}} p(\hat{G},G)\log \left(\frac{p(\hat{G}|G)}{p(\hat{G})}\right)d\hat{G}dG\\
    &\leq \int_G \int_{\hat{G}} p(\hat{G},G)\log \left(\frac{p(\hat{G}|G)}{r(\hat{G})}\right)d\hat{G}dG.
\end{split}
\end{equation}

Thus, the upper bound of \eqref{eq6} is as follows.
\begin{equation}
\begin{split}
    -I(\hat{G};T)+\beta I(\hat{G};G) &\approx H(\hat{Y}|Y)+\beta I(\hat{G};G) \\ &\leq H(\hat{Y}|Y)+\int_G \int_{\hat{G}} p(\hat{G},G)\log \left(\frac{p(\hat{G}|G)}{r(\hat{G})}\right)d\hat{G}dG \\
    &\leq H(\hat{Y}|Y)+\int_G \int_{\hat{G}}\log \left(\frac{p(\hat{G}|G)}{r(\hat{G})}\right)d\hat{G}dG \\
    &\approx H(\hat{Y}|Y)+\frac{\beta}{N}\sum_{i=1}^{N}\left[\log\left(\frac{p(\hat{G}_i|G_i)}{r(\hat{G})}\right) \right],
\end{split}
\end{equation}
where knowledge $G$ and $\hat{G}$ can be approximated with training samples.

\subsection{Mathematical Derivation in Section 5.3}
\begin{equation}
\begin{split}
    \hat{G} &\leq H(\hat{Y}|Y)+\frac{\beta}{N}\sum_{i=1}^{N}\left[\log\left(\frac{p(\hat{G}_i|G_i)}{r(\hat{G})}\right) \right] \\
    &\leq H(\hat{Y}|Y)+\frac{\beta}{N}\sum_{i=1}^{N}\left[\log\left(\frac{1}{r(\hat{G})}\right) \right]\\
    &=H(\hat{Y}|Y)+\frac{\beta}{N}\sum_{i=1}^{N}-\log r(\hat{G})\\
    &=H(\hat{Y}|Y)-\beta\log r(\hat{G}).
\end{split}
\end{equation}

\section{Gaussian Approximation}
In deep learning, we often need to solve for functional optimal. It is hard to do functional analysis in training models. The common method is to use differential (variational) approximation. Assuming that the unknown function obeys the Gaussian distribution, the analysis of the function is converted to the solution of the mean and variance of the Gaussian distribution. This process is called Gaussian approximation.
\begin{equation}
    q(x)=\mathcal{N}(x;\mu,\sigma^2),
\end{equation}
where $q(x)$ is the unknown function, $\mathcal{N}$ is Gaussian distribution, $\mu$ and $\sigma^2$ are mean and variance of $\mathcal{N}$.

The IAK's initialization in this paper is normalized Gaussian initialization, and we use Gaussian approximation to represent the parameter distribution of the trained IAK. This allows the functional analysis problem to be transformed into a two-unknowns function optimization problem, which can be calculated simply by using gradient descent method. 

\section{Experimental Settings}
All models in this paper are implemented
with Tensorflow 1.12 in Python 2.7 environment. All models are
trained with a cheif-worker distributed framework with 1600 CPUs.
To ensure fair comparison, all models are equipped with the same
embedding table, short-term and long-term sequence modules, and
debias module. The initial learning rate is set to 0.005 and the
optimizer is set to AdagradDecay. Besides, batch size is set
to 1024 and training epoch is set to 1. The activation function is set to LeakyReLU. All the DNNs in all models have three layers with the parameter set of [1024, 512, 256]. The embedding size is set to 8 for all categorical features. The dataset is split into training set and test set based on chronological order and the ratio is 6:1. The weights of loss functions are all set to 1 in all multi-task learning models. The encoding dimension is set to [10, 30, 50, 80, 100]. All Attention modules used are multi-head attention with 8 heads. There are two experts in MMoE. 

For the online A/B test, the users in two groups strictly satisfy the statistical assumption and the users' feedback actually happened.

\end{document}